\documentclass[aps,prd,twocolumn,showpacs,floats,floatfix,letterpaper,nofootinbib,superscriptaddress,]{revtex4-1}
\pdfoutput=1
\usepackage{hyperref}
\usepackage{adjustbox}
\usepackage{amssymb,amsmath,latexsym,mathrsfs}
\usepackage{graphicx,subfigure}
\usepackage{epsfig}
\usepackage{varioref,xr-hyper}
\usepackage{color}
\usepackage{multirow}
\usepackage{array}
\usepackage{placeins}
\hypersetup{
	colorlinks   = true,
	citecolor    = blue,
	linkcolor = red
}

\begin{document}
	
\title{Quantifying effects of inhomogeneities and curvature on gravitational wave standard siren measurements of $H(z)$}
\author{S. M. Koksbang}
\email{koksbang@cp3.sdu.dk}
\affiliation{$CP^3$-Origins, University of Southern Denmark, Campusvej 55, DK-5230 Odense M, Denmark}
	
\begin{abstract}
\noindent
For a flat $\Lambda$CDM universe, the dipole of the luminosity distance can be utilized to measure the Hubble parameter. It is here shown that this is not the case in more general settings where curvature and cosmic backreaction is permitted. This implies that a discordance between $H(z)$ measurements obtained using such dipole luminosity distance data and ``true''/actual $H(z)$ data obtained from e.g. cosmic chronometers is a signal of curvature and/or cosmic backreaction.
\newline\indent
By considering mock future gravitational wave measurements of the Hubble parameter obtained through the dipole luminosity distance, it is shown that already a $1\%$ curvature could in principle just barely show up in the determination. However, for realistic mock data generation using models with as much as 5 \% curvature, parameter estimates do not yield reliable measures of inconsistency between the false $H(z)$ measurements and true measurements of $H(z)$. At the same time, cosmic backreaction is hard to detect even if it makes up $10\%$ of the ``energy budget'' in the current universe, even when considering a highly idealized situation with low errors. The results concerning backreaction are based on specific  ``scaling solutions'' to the backreaction problem and the study shows that the possibility of detecting a signal of backreaction through the dipole of the luminosity distance depends strongly on the particular backreaction model.
\end{abstract}	
	
\maketitle
\flushbottom
\section{Introduction}
Direct measurements of the Hubble parameter, $H(z)$, are a valuable tool in cosmology, not the least because $H(z)$ depends strongly on the dark energy density and hence on dark energy phenomenology, but also e.g. because measurements of $H(z)$ may hold key insights into the $H_0$-tension \cite{tension1,tension3,tesnion2} as studied in e.g. \cite{Hubble_tenion, Hubble_tension_H, H0_H_added} (see e.g. the introduction of \cite{tension_summary} for a summary of the Hubble-tension and possible solutions). Currently, the Hubble parameter has been measured directly by using baryon acoustic oscillations (BAO) \cite{BAO1,BAO2,BAO3, BAO4} and cosmic chronometers \cite{cc0,cc_theory1,cc_theory2,cc_theory3} data and the data is usually combined when used for constraining model parameters. The measurements of especially the latter type are still in their infancy and more data as well as a better understanding of uncertainties and systematics are needed for the probes to become useful for a precise determination of $H(z)$. Other future direct probes of the Hubble parameter are also known, such as redshift drift measurements \cite{sandage,mcVittie} and measurements of the dipole of the luminosity distance \cite{dipole}.
\newline\indent
These probes have all mainly been developed within the framework of Friedmann-Lemaitre-Robertson-Walker (FLRW) models which describe universes that are exactly spatially homogeneous and isotropic. However, the real universe has structures which can affect the large scale ``average'' evolution of the Universe through cosmic backreaction \cite{fluid1, perfect, general} (see e.g. \cite{review1,review2,review3} for reviews and \cite{further_bc,further_bc2} for further considerations on generalizations of the original scheme). It is not clear to what extent backreaction actually affects the dynamics of our universe -- while it may be entirely negligible, it is also possible that backreaction is important at percent-level and it may even lead to accelerated expansion and hence mimic dark energy \cite{conjecture1,conjecture2,conjecture3,conjecture4}.
\newline\indent
As discussed in \cite{multipleH}, multiple probes of the Hubble parameter may help us understand what role backreaction plays in our universe because the different probes actually measure different things if we are not in a universe that behaves as an FLRW universe on large scales. In particular, the results of \cite{2region,Hellaby,Asta_dz} imply that redshift drift determinations of the Hubble rate will not measure the volume averaged expansion rate in a universe with non-negligible backreaction (or any other significant deviation of FLRW geometry which has been illustrated using various models \cite{dz_szekeres1,dz_szekeres2,stephani,bianchi}). Indeed, as pointed out in \cite{multipleH}, the only currently known probe of the Hubble parameter which clearly yields a measurement of the actual volume averaged/large scale expansion rate in a spatially statistically homogeneous and isotropic spacetime with non-negligible backreaction is that based on cosmic chronometers. It does not seem likely that the dipole luminosity distance and BAO measurements of the Hubble parameter actually measure the volume averaged expansion rate but a definite answer is not given by the current literature. The work presented here seeks to partially remedy this by studying the dipole luminosity distance determination of the Hubble parameter.
\newline\indent 
In \cite{dipole} it was shown that under certain conditions, the dipole of the luminosity distance can be used to directly measure the Hubble parameter. Specifically, if the Universe is well-described by a perturbed flat FLRW model, then the monopole luminosity distance is the well-known relation
\begin{align}
	d_L^0(z) = (1+z)\int_0 ^z\frac{dz'}{H(z')}
\end{align}
and the main contribution to the dipole will be due to our peculiar motion. In that case, the dipole is given by
\begin{align}\label{eq:dipole_flat}
	d_L^1(z) = \frac{|v_0|(1+z)^2}{H(z)},
\end{align}
where $v_0$ is the peculiar velocity of the observer relative to a cosmological reference frame. Assuming that $v_0$ is the main source of the dipole seen in the CMB as well as in the mean luminosity distance, we can infer $|v_0|$ from the CMB dipole. In that case, the dipole of the luminosity distance directly yields the Hubble parameter. Although it is usually assumed that our peculiar velocity can be deduced from the dipole of the CMB, it should be noted that there are some indications in observational data that this expectation is violated \cite{anisotropy1,anisotropy2,anisotropy3}. Nonetheless, it will here be assumed that $|v_0|$ can be extracted from other data than that used for obtaining the luminosity distance.
\newline\indent
While \cite{dipole} considered supernova data, it was in \cite{dipole_GW} noted that a percent-level determination of $H(z)$ would require $\sim 10^6$ supernova observations, which is not realistic. Gravitational waves can be used as standard sirens though \cite{GW_sirenes,sirene1} and have already been used for measuring the Hubble constant \cite{GW_H0,GW_H0_2} (see e.g. also \cite{Mukherjee:2022afz} for the latest constraint of $H_0$ based on cross-correlation of gravitational wave sources with galaxies \cite{Mukherjee:2020hyn, Diaz:2021pem}). With the expected high number of detected gravitational wave signals in the near future, luminosity distance measurements based on these standard sirens are expected to yield estimates of the Hubble parameter at percent precision \cite{dipole_GW,dipole_GW2}. However, if the Universe is not well-described by a flat FLRW spacetime on large scales, the expression in equation \ref{eq:dipole_flat} is not correct. A comparison of ``$H(z)$'' measurements obtained from standard siren dipole luminosity distances with true measurements based on cosmic chronometers, should therefore reveal a discordance between the measurements. Moreover, the expression used for obtaining $H(z)$ through measurements of the dipole of the luminosity distance is only correct for a {\em flat} FLRW universe. If we live in a curved universe, erroneously using equation \ref{eq:dipole_flat} will lead to $H(z)$ measurements in discordance with $H(z)$ measurements from e.g. cosmic chronometers and future redshift drift measurements. This discordance can e.g. be measured through the ``index of consistency'', IOI, defined in \cite{discordance}. 
\newline\indent
In the following sections, the expression for the dipole component of the luminosity distance is studied for general spacetimes with spatial statistical homogeneity and isotropy. After presenting the theoretical considerations, a numerical investigation is used to estimate how large deviations from flat FLRW are required for obtaining a signal in the $H(z)$ measurement. The investigation is first done qualitatively, where-after a quantitative analysis is conducted with the final goal being computing the IOI.

\section{The dipole luminosity distance beyond flat FLRW spacetimes}
This section serves to present the computation of the dipole of the luminosity distance in a general spacetime exhibiting statistical spatial homogeneity and isotropy.
\newline\newline
If the considered spacetime is only statistically and not exactly spatially homogeneous and isotropic, even a Copernican (i.e. random) observer will see some directional dependence of the redshift-distance relation. Although the spacetimes considered here are not necessarily close to a specific FLRW model, it will be assumed that the statistical homogeneity and isotropy still leads the observed dipole of the luminosity distance to  be induced by the peculiar motion of the observer, as it was shown to be the case in standard perturbed FLRW spacetimes in \cite{Bonvin1}. In that case, the expression for the dipole can be obtained using a derivation similar to that used in \cite{dipole, dipole_GW} for the flat FLRW case.
\newline\indent
In order to most easily take the peculiar velocity of the observer into account, it is prudent to start by considering the angular diameter distance which is defined as
\begin{align}
	d_A^2 :=\frac{\delta A}{\delta \Omega},
\end{align}
where $\delta A$ is the area of the light beam at the emitter and $\delta\Omega$ is the solid angle of the beam. For an observer with peculiar velocity $|v_0|<<1$ (setting $c=1$) on the spatial hypersurfaces of statistical homogeneity and isotropy, the observed angular diameter distance will (through the special relativistic transformation of $\delta\Omega$) be modified by a factor of $(1+n_{i}v_0^{i})$, where $n^{i}$ is the direction of observation. In addition, the redshift is modified by a Doppler term according to (first order in $v^i$)
\begin{align}
	z_{\rm obs} = z_{\rm monopole} -(1+ z_{\rm monopole})\cdot n_{i}v_0^{i},
\end{align}
where $z_{\rm obs}$ is the observed redshift and $z_{\rm monopole}$ is the redshift which would be observed by an observer comoving on the hypersurfaces of statistical homogeneity and isotropy.
\newline\indent
Introducing the luminosity distance by the distance duality relation and combining with the above considerations and a Taylor expansion of $d_L$ around $z_{\rm monopole}$, the luminosity distance is seen to be given by
\begin{align}\label{eq:dipole}
\begin{split}
d_L(z) = d_L^0(z) + \left( (1+z)\frac{d}{dz}d_L^0(z) -d_L^0(z) \right) \cdot n_iv_0^i
\end{split}
\end{align}
For an FLRW metric, this implies
\begin{align}\label{eq:dipole_FLRW}
d_L^1(z) = \frac{|v_0|(1+z)^2}{H(z)}\cdot \left\{ \begin{array}{rl}
  \cos\left(R_0^{-1}\int_0^z\frac{dz'}{H(z')} \right) &\text{if} \,\, K = 1 \\
1 &\text{if}\,\, K = 0\\
  \cosh\left(R_0^{-1}\int_0^z\frac{dz'}{H(z')} \right) &\text{if} \,\, K = -1
\end{array} \right.,
\end{align}
where $R_0$ is the curvature radius and
\begin{align}
d_L^1/|v_0|:=(1+z)\frac{d}{dz}d_L^0(z) -d_L^0(z)
\end{align}
is the size of the dipole contribution to the luminosity distance. 
\newline\newline
While equation \ref{eq:dipole_FLRW} requires the monopole contribution of the luminosity distance to be given by that of an FLRW universe, the derivation up to and including equation \ref{eq:dipole} is more general. Equation \ref{eq:dipole} is valid under the assumptions that the universe is spatially statistically homogeneous and isotropic for a random observer with peculiar velocity $v_0$ in the hypersurfaces of statistical homogeneity and isotropy. It is also assumed that structure evolution is slow compared to the time it takes a light ray to traverse the homogeneity scale so that effects of inhomogeneity on $d_L$ and $z$ other than those due to the peculiar velocity of the observer can be ignored when considering mean observations. Under these same assumptions, the mean redshift-distance relation -- i.e. the redshift-distance relation obtained by averaging over many random lines of sight -- is given by \cite{light1,light2}
\begin{align}\label{eq:DAsyksy}
	\frac{d^2}{dz^2}d^0_A = -\frac{4\pi G\rho_D}{\left( (1+z)H_D\right) ^2}d^0_A - \frac{d}{dz}d^0_A \left(\frac{2}{1+z} + \frac{dH_D}{dz}H_D \right) ,
\end{align}
where $H_D$ is a third of the spatially averaged expansion rate and $\rho_D$ is the spatially averaged energy density of the content of the universe. Spatial averages are defined according to Buchert's averaging scheme \cite{fluid1} such that
\footnote{This average is strictly only valid when assuming a dust content of the Universe comoving on those hypersurfaces which are here taken to be those of statistical homogeneity and isotropy. Here the interest is specifically observers with a non-comoving velocity field. The physical picture is, however, similar to what is assumed in standard cosmology based on FLRW backgrounds: While the local observer may have peculiar velocity, it is assumed that the content of the Universe can be described as a comoving dust on the hypersurfaces of statistical homogeneity and isotropy on sufficiently large scales, still small enough for averaging over inhomogeneities to be relevant. See e.g. also \cite{light2} for considerations of how to include peculiar velocity of the observer.} 
\begin{align}
	\left\langle x \right\rangle _D = x_D :=\frac{\int_Dx\sqrt{|\det g^{(3)}|}}{\int_D\sqrt{|\det g^{(3)}|}}
\end{align}
for a scalar $x$ and where the spatial domain $D$ is assumed larger than the homogeneity scale in order for the average quantities to describe the large scale universe. In order to connect spatial averages with observations as done with the angular diameter distance above, it is crucial that spatial averages are computed on the spatial hypersurfaces of statistical homogeneity and isotropy.
\newline\indent
It was also argued in \cite{light1,light2} that the relation between the redshift and the volume scale factor, $a_D$, upon averaging over many lines of sight is
\begin{align}\label{eq:zSyksy}
	1+z_D = \frac{1}{a_D},
\end{align}
where $a_D$ is defined through the proper volume, $V_D$, of the domain $D$ according to $a_D:=\left(V_D/V_{D,0} \right) ^{1/3}$. The relations \ref{eq:DAsyksy} and \ref{eq:zSyksy} have been tested in various inhomogeneous cosmological models and generally seem robust as long as opaque regions are not introduced and the above mentioned assumptions of statistical homogeneity and isotropy and slowly evolving inhomogeneities are fulfilled \cite{2region,Hellaby, DR, towards}.
\newline\indent
As discussed in \cite{light1}, equation \ref{eq:DAsyksy} does not generally reduce to a simple integral expression - this only happens in the FLRW limit. As the superscript on $d_A^0$ in equation \ref{eq:DAsyksy} indicates, the angular diameter distance obtained from that equation corresponds to the monopole contribution. Solving the equation and applying the distance-duality relation thus gives $z, d_L^0, dd_L^0/dz$ from which the dipole can be computed using equation \ref{eq:dipole}. Although it is not immediately apparent how the resulting dipole will depend on $H_D$, it {\em is} apparent that the relation will not be as simple as for the flat FLRW models. Indeed, already for non-flat FLRW models, the relation becomes more complicated. Thus, if we do not live in a universe which is well described by a single flat FLRW model on large scales, using equation \ref{eq:dipole_flat} on luminosity distance data will not actually yield a measure of the large scale expansion rate. This will be illustrated below for models with modest curvature and modest signatures of inhomogeneities through cosmic backreaction. Before this, the concept of cosmic backreaction will be elaborated in the next section together with an introduction to specific models of cosmic backreaction.

\section{Models of cosmic backreaction}
By employing Buchert's averaging scheme on Raychaudhuri's equation and the Hamiltonian constraint, the equations (setting $G = 1$)
\begin{align}\label{eq:Friedmann_like}
3H_D{}^2:=\frac{1}{3}\Theta_D{}^2 = 3\left(\frac{a_{D,t}}{a_D} \right) ^2 = 8\pi \rho_D-\frac{1}{2}R_D+\Lambda-\frac{1}{2}Q
\end{align}
and
\begin{align}\label{eq:acc_like}
3\frac{a_{D,tt}}{a_D} = -4\pi \rho_D+ \Lambda + Q,
\end{align}
are obtained. $\Theta$ is the local dust expansion rate and $\sigma^2:=\frac{1}{2}\sigma_{\mu\nu}\sigma^{\mu\nu}$ is the shear scalar of the dust. The term $Q:=\frac{2}{3}\left(\left\langle\Theta^2 \right\rangle_D -\left\langle \Theta\right\rangle_D{} ^2 \right)-2\left\langle \sigma^2\right\rangle_D  $ is the kinematical backreaction. The two equations above describe the large-scale/average evolution of an inhomogeneous universe. The equations are reminiscent of the Friedmann equations but the kinematical backreaction drives the large scale evolution away from FLRW dynamics. When $Q$ is non-vanishing, it is coupled to the spatially averade spatial Ricci scalar $R_D$ according to the integrability condition
\begin{align}
a_D{}^{-6}\partial_t\left(a_D{}^6Q \right)+a_D{}^{-2}\partial_t\left( a_D{}^2R_D\right) =0
\end{align}
which ensures consistency between the two previous equations. When the kinematical backreaction vanishes identically, the averaged Ricci scalar is proportional to $a_D{}^{-2}$ and the averaged dynamical equations reduce to the Friedmann equations.
\newline\indent
The above set of equations does not form a closed set. Families of solutions can nonetheless be found by introducing ansatzes on e.g. the form of $Q$ and $R_D$. One simple family of solutions is the scaling solutions of \cite{scaling}. The scaling solutions is a family of solutions to the integrability condition given by
\begin{align}
\begin{split}
R_D &= R_{D_0}a_D{}^{n_D}\\
Q&= -\frac{n_D+2}{n_D+6}R_D,
\end{split}
\end{align}
where $n_D\neq -6, -2$.
\newline\indent
It should be stressed that the scaling solutions are not {\em a priori} based on physical justifications but rather on mathematical simplicity. However, the particular solution with $n_D = -1$ has some limited physical motivation since various types of perturbation theory \cite{1overa2,1overa3} indicate this as the dominant behavior of $Q$. In the study based on numerical relativity in \cite{1overa1} it was found that $n_D = -1$ for small initial perturbations around an FLRW background, while $n_D = 1$ when larger initial perturbations are used. This latter result indicates that the perturbative result yielding $n_D = -1$ should probably not be taken too seriously and that if one insists on considering $n_D = -1$ physically motivated, then $n_D = 1$ should perhaps be considered equally motivated.
\newline\indent
For the goal of the following section, concrete models for $Q$ and $R_D$ are needed. Thus, for the lack of better options, $n_D = \pm 1$ will be considered in order to quantify the size of backreaction necessary for generating a signal through the dipole determination of the luminosity distance. By considering two values of $n_D$ rather than just one, the importance of the specific form versus the numerical size of $Q$ can be estimated and hence used to evaluate the reliability of the obtained results.
\newline\newline
For quantifying the ``size'' of backreaction in the considered models, it is convenient to introduce the density parameters $\Omega_m:=\frac{8\pi}{3H_{D_0}{}^2}\rho_{D_0}$, $\Omega_{\Lambda}:=\frac{\Lambda}{3H_{D_0}{}^2}$, $\Omega_R:=\frac{-R_{D_0}}{6H_{D_0}{}^2}$ and $\Omega_Q:=-\frac{Q_0}{6H_{D_0}{}^2}$. With these, the averaged Hamiltonian constraint can be written as (note that $\rho_D\propto a_D{}^{-3}$ \cite{fluid1})
\begin{align}\label{eq:Hubble_short}
\frac{H_D{}^2}{H_{D_0}{}^2}=\Omega_m a_D{}^{-3}+\Omega_{\Lambda}+\frac{4}{n_D+6}\Omega_R a_D{}^{n_D}
\end{align}
which evaluated at present time simply reads 
\begin{align}\label{eq:one}
1=\Omega_m+\Omega_{\Lambda}+\frac{4}{n_D+6}\Omega_R .
\end{align}
Assuming that all curvature stems from $Q$ being non-zero, the size of backreaction will then be used to mean the size of $\frac{4}{n_D+6}\Omega_R$ which indicates the percentile of the present time ``energy budget'' that can be attributed to cosmic backreaction.

\section{Numerical investigation}
In this section, the luminosity distance including its dipole will be computed for different non-flat and/or non-FLRW models. This will then be used together with equation \ref{eq:dipole_flat} to extract measurements of ``$H(z)$''. Since the considered models are not actually flat FLRW models, the extracted $H(z)$ will not be equal to the true $H(z)$. In other words, there is a discordance between $H(z)$ extracted from the dipole luminosity distance measurements and the true $H(z)$.
\newline\indent
The first subsection below serves to present an initial qualitative investigation of the expected discordance. Afterwards, a quantitative investigation with mock $H(z)$ data is presented.

\subsection{Qualitative investigation}\label{subsec_qualitative}
\begin{figure}
\centering
\includegraphics[scale = 0.5]{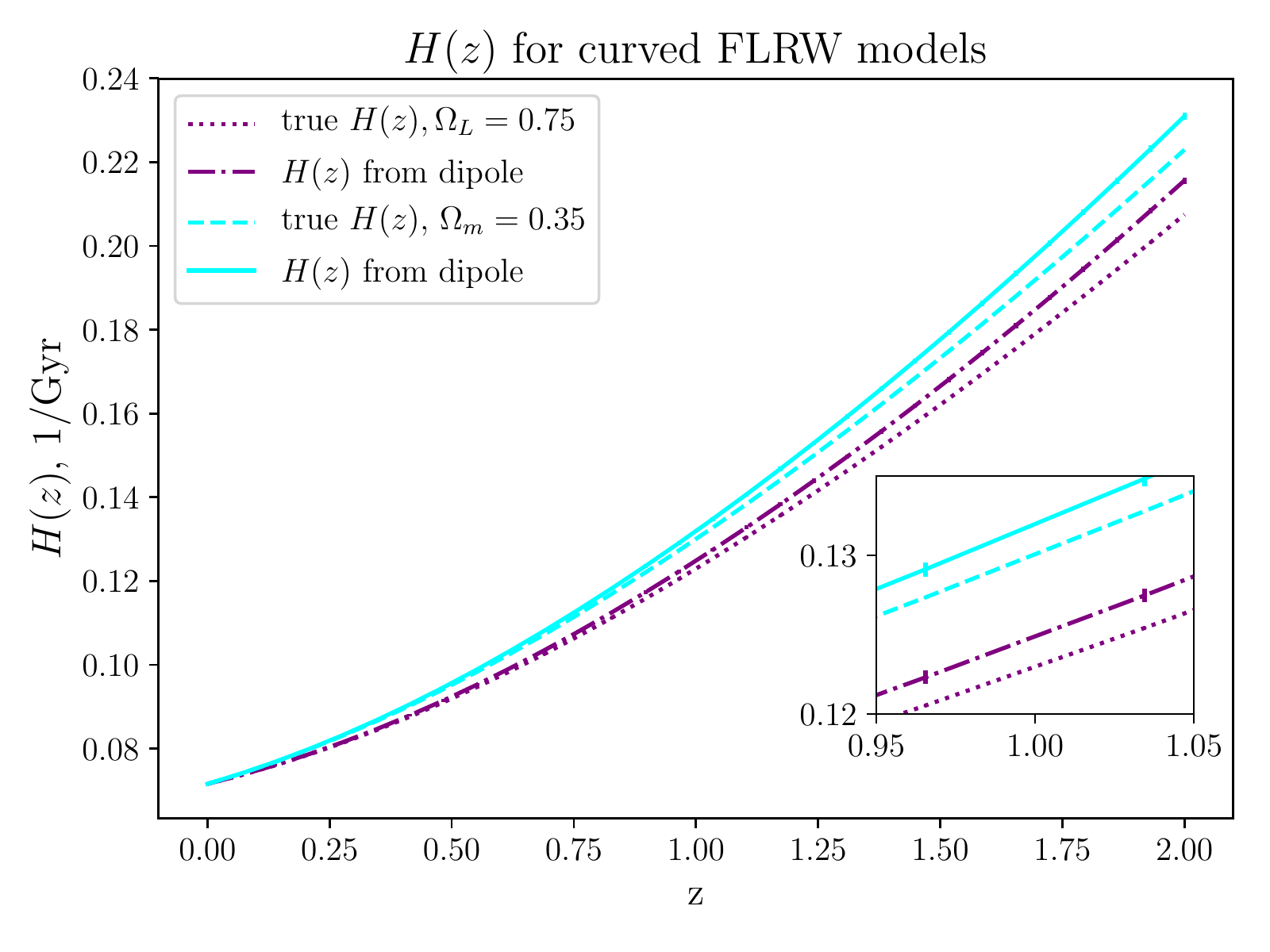}
\caption{Hubble parameter for curved FLRW models with curvature induced by increasing either $\Omega_m$ or $\Omega_\Lambda$ as indicated in the figure labels. Lines labeled ``$H(z)$ from dipole'' represent the measured quantity when combining the measured luminosity distance dipole with the expression for the dipole in flat FLRW models. Error bars represent 1.5 \% errors on the measurement.}
\label{fig:curvature}
\end{figure}

\begin{figure}
\centering
\includegraphics[scale = 0.5]{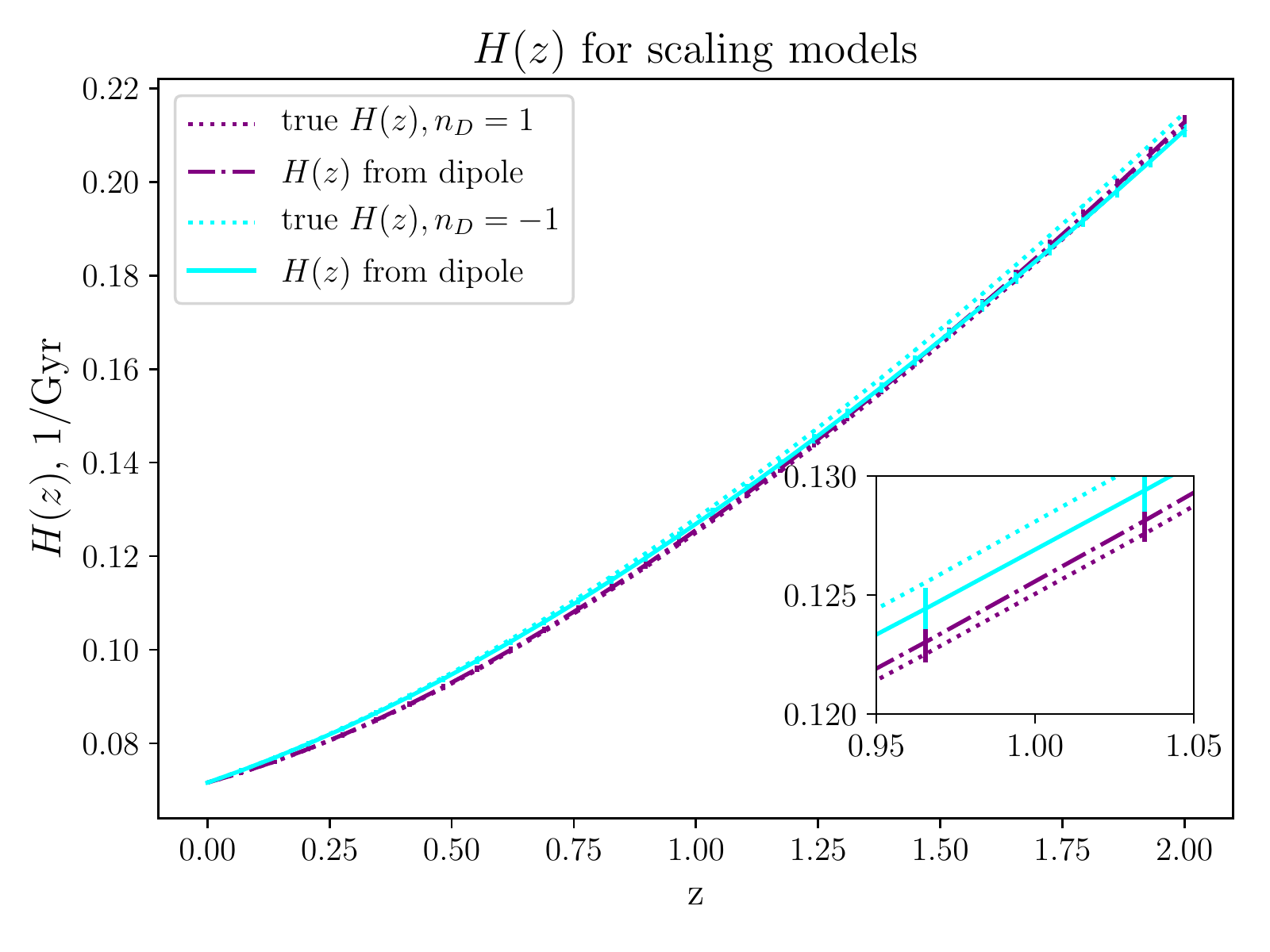}
\caption{Hubble parameter for scaling models with backreaction introduced through equation \ref{eq:one} by decreasing $\Omega_{\Lambda}$ to 0.6 with fixed $H_0$. Labels indicate whether $n_D = \pm 1$ was used. Lines labeled ``$H(z)$ from dipole'' represent the measured quantity when combining the measured luminosity distance dipole with the expression for the dipole in flat FLRW models. Error bars represent 1.5 \% errors on the measurement. The lines lie nearly on top of each other and can only really be distinguished in the close-up.}
\label{fig:backreaction}
\end{figure}

In this section, the size of the difference between the inferred and actual $H(z)$ is compared with expected uncertainties of the measurements. This uncertainty is estimated based on the computations of \cite{dipole_GW}, where the error in $H(z)$ measurements up to $z = 1$ were found to be 0.75 - 1.5 \% for 10 years of observation with the proposed Japanese gravitational wave observatory DECIGO \cite{DECIGO}. At $z = 2$ the error has increased to $6\%$. These estimates ignore errors from lensing which can significantly affect the precision of the total error as shown in \cite{dipole_GW2}. Lensing will also be ignored here though, under the expectation that it will be possible to remove systematics from lensing in the near-future (errors from lensing will be considered in the next subsection). Note also that the error estimates of \cite{dipole_GW,dipole_GW2} are in principle model-dependent. This model-dependence is neglected throughout under the expectation that the order of magnitude would be the same for an estimate based on the particular individual models -- this is justified by the fact that the studied models aim at considering only {\em small} deviations from flat FLRW. Lastly, note that although the errors used here are specifically for DECIGO, the errors are expected to be the same order of magnitude for LISA and the Einstein telescope \cite{same_LISAET}.
\newline\indent
The results are shown using $|v_0| = 369.8\rm km/s$ as inferred with Planck \cite{cmb}.
\newline\newline
Although the common expectation is that the Universe is flat, some studies indicate that the universe may be slightly curved \cite{curvature1, curvature2}. It is therefore interesting to see if a small amount of curvature will appear as a measurable deviation between the actual large scale expansion rate and that inferred by erroneously using equation \ref{eq:dipole_flat}. Figure \ref{fig:curvature} shows $H(z)$ measurements in the redshift interval $z\in [0,2]$ for FLRW models with small amounts of curvature generated by either increasing $\Omega_{m}$ or $\Omega_{\Lambda}$ compared to the default values of $\Omega_m = 0.3$ and $\Omega_\Lambda = 0.7$. $H_0 = 70\rm km/s/Mpc$ is used throughout. The results are shown for a 5 \% curvature. With this curvature, the precision is good enough to detect a difference between the actual $H(z)$ and that inferred by wrongly employing equation \ref{eq:dipole_flat} (the expression for the dipole of the luminosity distance in a flat FLRW model) to the data. Note that $1.5\%$ errors are used all the way to $z = 2$ even though the error at this value of the redshift is expected to have increased to around 6 \%. A $6\%$ error is large enough to hide the difference between the true $H(z)$ and the one inferred from the measurement. However, even with the high estimate of the error in the low-z ($z<1$) region, it becomes possible to identify the curvature if it is around 5 \%. 
\newline\indent
Figure \ref{fig:curvature} only shows the results for two particular ways of obtaining a positive 5 \% curvature. The results do not depend strongly on this though. It is also notable that a curvature of 1 \% (not shown) is just on the verge of being detectable with a precision of $0.75\%$ at $z = 1$.
\newline\newline
Figure \ref{fig:backreaction} shows $H(z)$ measurements obtained by applying equation \ref{eq:dipole_flat} to data in a universe with non-negligible cosmic backreaction. The figure shows results for $n_D =\pm 1$ and with $\Omega_{R}$ determined from equation \ref{eq:one} by setting $\Omega_m = 0.3$, $\Omega_\Lambda = 0.6$ and with $H_0 = 70$km/s/Mpc. The choice of reducing $\Omega_\Lambda$ from $0.7$ to $0.6$ rather than, say, reducing $\Omega_m$ is made based on cosmic backreaction mainly being discussed in terms of its ability to mimic a dark energy component.
\newline\indent
For the two considered models, the difference between the inferred and actual large scale Hubble parameter is very modest and can only for the $n_D = -1$ case barely be detected with the $1.5\%$ error at $z = 1$. While this result should be evaluated remembering that $1.5\%$ is the the higher bound on the estimated precision, it must also be remembered that the ``size'' of the backreaction contribution is at 10 \% in these models and thus quite large.
\newline\indent
It is also noteworthy that the close-up in figure \ref{fig:backreaction} shows a quite significant difference in the deviation between true and inferred $H(z)$ depending on the two choices $n_D = \pm 1$. Since there is only little physical justification for these particular scaling models (and none for the scaling models in general), the significant difference between the results of the two models means that the results cannot be trusted in terms of quantifying to what extent backreaction will lead to signatures in the determination of the Hubble parameter using the dipole of the luminosity distance. The significance of the signature will depend on the specific backreaction model and must therefore be computed separately for any given model one wishes to study.
\newline\newline
The qualitative investigation of this section shows that the possibility of revealing cosmic backreaction by comparing $H(z)$ measurements obtained from the luminosity distance with true values of $H(z)$ depends strongly on the particular backreaction parameterization. In addition, for the models used here, it is clear that a significant amount of backreaction -- of order $10\%$ -- is necessary for the resulting signal to be measurable even using ideal data as above. With significant backreaction, it is reasonable to speculate that effects of inhomogeneity will affect gravitational wave propagation in ways not taken into account here. Indeed, this topic was recently studied in \cite{inhomogeneity_GW} where it was found that backreaction affects gravitational wave observables. Therefore, the quantitative analysis presented below is restricted to curved FLRW models.

\subsection{Quantitative investigation of data in a curved FLRW universe}
As shown above, the $H(z)$ data inferred from observational data obtained with standard sirens does not actually represent $H(z)$. On the other hand, as shown in \cite{multipleH}, cosmic chronometers data does indeed yield $H(z)$. In addition, in a curved FLRW model redshift drift will also yield true measurements of $H(z)$. While redshift drift measurements are expected achieved within a few decades (see e.g. \cite{dz_decade, dz_decade2,dz_decade3,dz_decade4,decade5}) cosmic chronometers data is already available but at this point the data has too large errors to be useful here (and this is despite the fact that recent investigations indicate that the errors on some cosmic chronometers data may be too optimistic \cite{too_optimistic} -- see e.g. also the discussion in section 7.11 of \cite{too_optimistic2}). However, since cosmic chronometers data is still in its infancy, it is reasonable to expect that the technique will yield much more precise estimates of $H(z)$ within a few decades. To be able to construct concrete mock data of the true $H(z)$ value (envisioned obtained with e.g. cosmic chronometers and redshift drift), the errors on this mock data will here be set equal to the error on the standard siren data.
\subsubsection{Generating mock data}\label{subsub:generating_mock}
Mock DECIGO data is constructed using the procedure presented in \cite{dipole_GW} and more recently also used in e.g. \cite{dipole_GW2}. Error estimates include a lensing contribution, $\sigma_{\rm lens}$, an instrumental contribution, $\sigma_{\rm inst}$, as well as an error induced by the peculiar velocities of host galaxies, $\sigma_{\rm pv}$. The instrumental error associated with DECIGO is detailed in \cite{dipole_GW}, while the other two errors are given by
\begin{align}
	\sigma_{\rm lens} = 0.066\cdot \left[ \frac{1-(1+z)^{-1/4}}{0.25} \right] ^{1.8}
\end{align}
and
\begin{align}
	\sigma_{\rm pv} = \left| 1-\frac{(1+z)^2}{H(z)d_L^0} \right| \sigma_{\rm gal},
\end{align}
where $\sigma_{\rm gal} = 300 \rm km/s$ is due to galaxy velocity dispersion and given in \cite{gal}.
\newline\indent
Combining the three errors yields an error estimate for the monopole measurement of the luminosity distance, 
\begin{align}
\left( 	\frac{\Delta d_L^0}{d_L^0}\right) ^2 = \sigma_{\rm lens}^2 + \sigma_{\rm inst}^2 + \sigma_{\rm pv}^2.
\end{align}
From this, the relative error on $H(z)$ inferred from the DECIGO standard siren dipole luminosity distance data is \cite{dipole, dipole_GW,dipole_GW2}
\begin{align}
	\frac{\Delta H}{H} = \sqrt{3}\frac{d_L^0}{d_L^1}\frac{\Delta d_L^0}{d_L^0}.
\end{align}
The redshift distribution of the mock data follows a probability (see e.g. \cite{dipole_GW2,distribution})
\begin{align}
	P\propto \frac{4\pi d_L^2R}{(1+z)^3H},
\end{align} 
where
\begin{align}
R = \left\{ \begin{array}{rl}
1+2z &\text{if} \,\, 0<z<1 \\
\frac{3}{4}(5-z) & \text{if} \,\, 1\leq z<5\\
0 & \text{otherwise}
\end{array} \right..
\end{align}
Using this distribution, mock data representing the expected \cite{DECIGO} number of $10^6$ events is produced by adding errors to ideal measurements based on a Gaussian random distribution with standard deviation equal to $\Delta H$. Afterwards, the measurements are binned into redshift bins of width 0.1.
\begin{figure*}
	\centering
	\subfigure{
		\includegraphics[scale = 0.5]{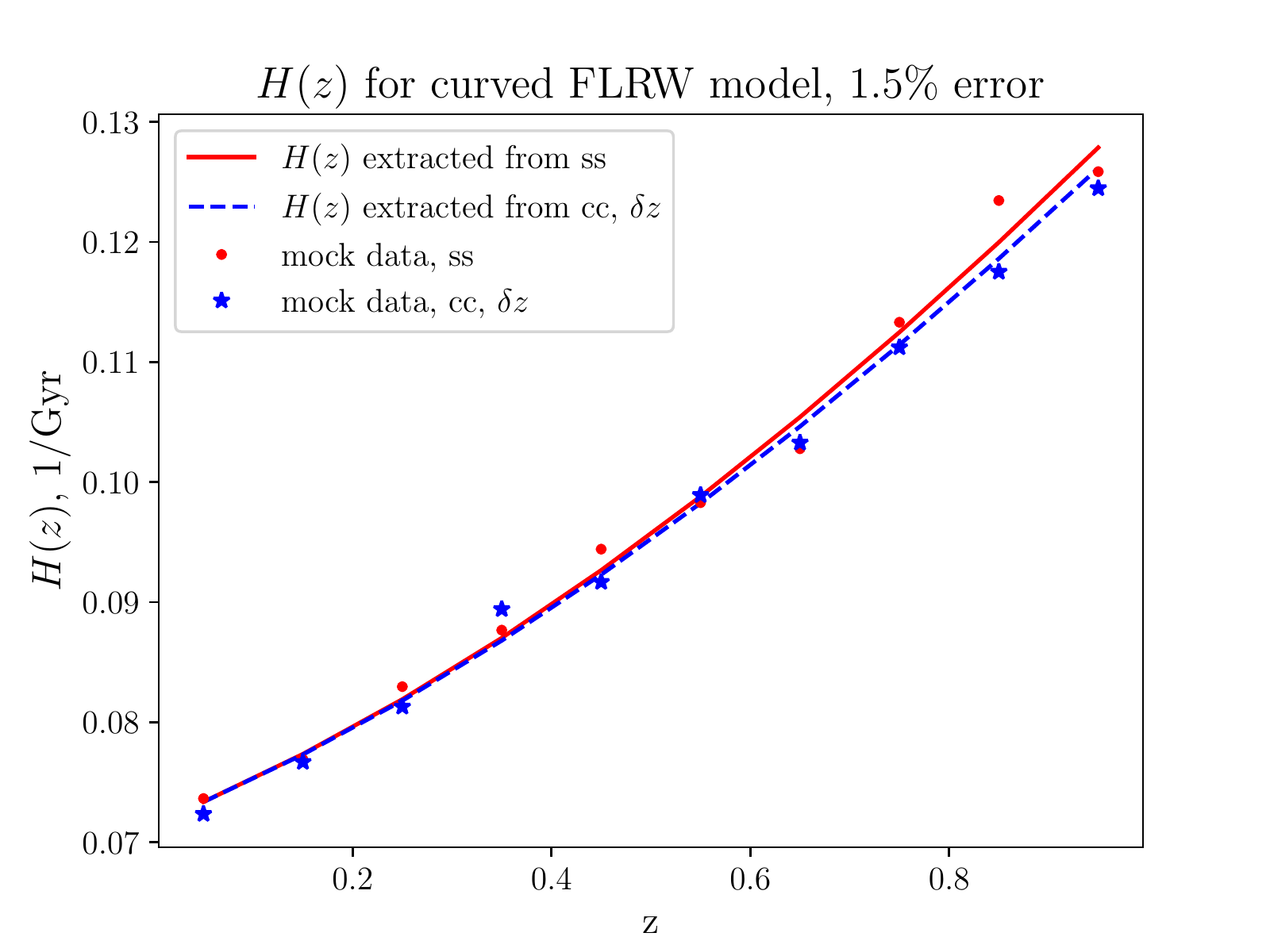}
	}
	\subfigure[]{
		\includegraphics[scale = 0.5]{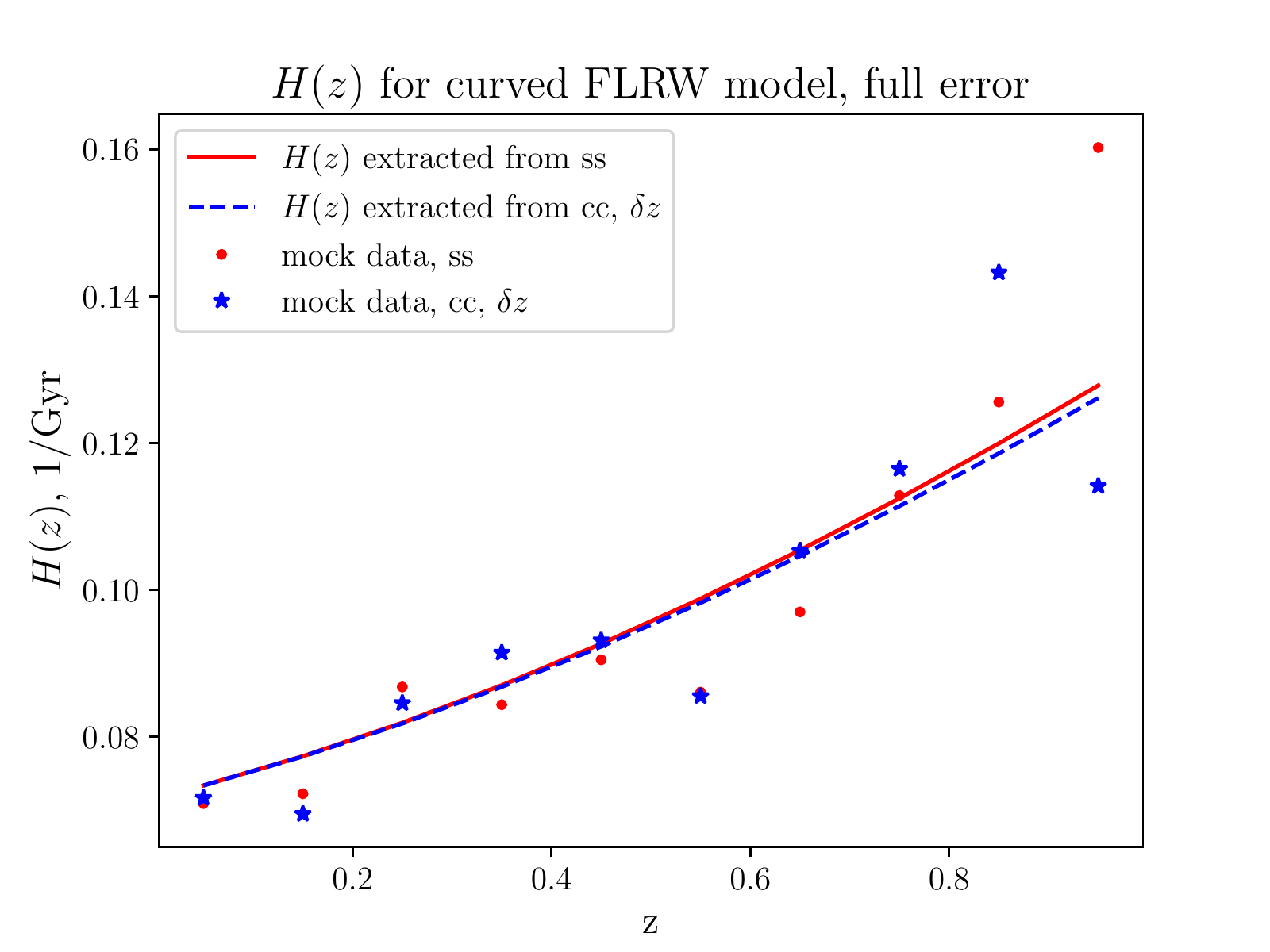}
	}
	\caption{(Binned) Mock DECIGO data using 1.5 \% and full error estimates. The mock DECIGO data (``ss'') is plotted together with mock $H(z)$ measurements based on cosmic chronometers  (``cc''), redshift drift (``$\delta z$'') or similar measurement yielding the true expansion rate measurement. The mock data is plotted together with curves representing the exact values -- in the case of the DECIGO data, computed by (wrongly) employing  equation \ref{eq:dipole_flat}. The data is based on a curved $\Lambda$CDM model with $H_0 = 70$km/s/Mpc, $\Omega_{m} = 0.35$ and $\Omega_{\Lambda} = 0.7$.}
	\label{fig:mock}
\end{figure*}
\subsubsection{Model fitting and cosmological discordance}
Figure \ref{fig:mock} shows two sets of mock data. The left figure has a low error with standard deviation of 1.5 \% to draw a parallel to the qualitative results of section \ref{subsec_qualitative}. The mock data presented in the figure to the right was instead generated using the full error described above. Data is only shown up to $z = 1$ and this is also the only data which is used in the following since the precision of data at higher redshift is not high enough to be useful for the analysis. When generating the data with 1.5\% error only 10 data points are used and these are evenly distributed in the redshift range $z\in[0,1]$ to mimic the binning necessary to reduce the error to this value.
\newline\indent
It is evident already from the plots of the mock data, that the mock data generated with the full error cannot possibly be used to reveal a discordance between the true $H(z)$ and that extracted by (wrongly) employing equation \ref{eq:dipole_flat} to the DECIGO data. Similarly, an inspection of figure \ref{fig:mock} makes it clear that it will be difficult to reveal a discordance between the data sets even with the low error. However, figure 3 does not represent the true situation; equation \ref{eq:dipole_flat} is assumed employed on the DECIGO data because cosmological data is mostly studied using the assumption that the Universe behaves as a flat $\Lambda$CDM universe. Therefore, the assumption here is that any parameter determinations based exclusively on the $H(z)$ measurements will base parameter estimates on the flat $\Lambda$CDM model. In a curved universe, this will also lead to a wrong inference of cosmological parameters from true $H(z)$ measurements. This point is not encapsulated in figure \ref{fig:mock}. The inferred parameter values should nonetheless be in agreement if comparing cosmic chronometers and redshift drift data -- although the parameter estimations will agree on wrong values. However, when comparing with parameter estimates based on dipole luminosity data, a discordance should show up -- if the data is precise enough.
\newline\indent
\begin{figure}
	\centering
	\includegraphics[scale = 0.5]{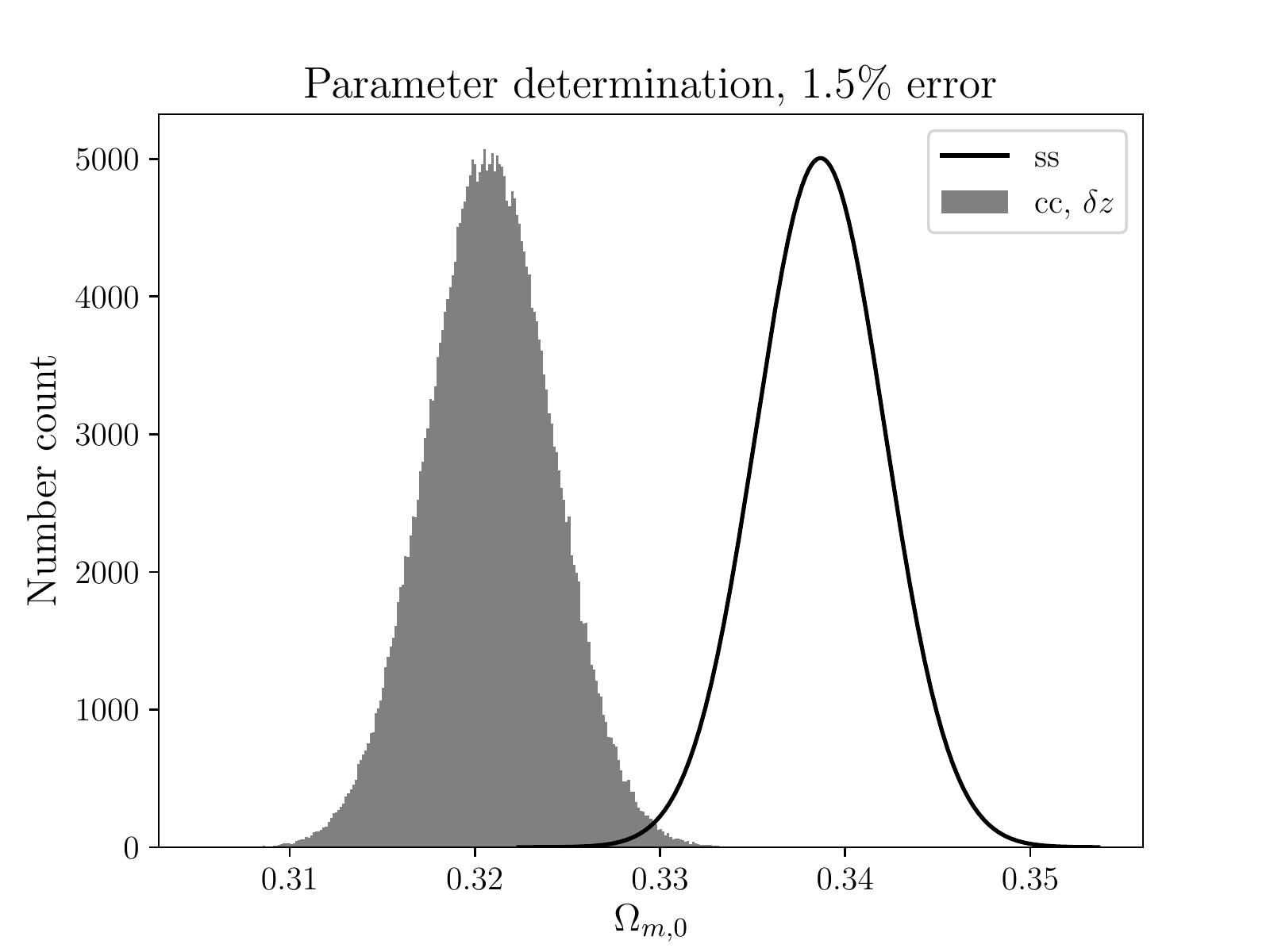}
	\caption{Distribution of $\Omega_{m}$ in MCMC chains obtained with emcee \cite{emcee} using 1.5\% error on mock data. The distribution marked ``ss'' represents results obtained with the standard siren data, i.e. by wrongfully employing equation \ref{eq:dipole_flat} on mock dipole luminosity distance data from DECIGO. The distribution marked ``cc, $\delta z$'' represents mock data yielding the true $H(z)$, such as cosmic chronometers or redshift drift data.}
	\label{fig:omega_15}
\end{figure}
\begin{figure}
	\centering
	\includegraphics[scale = 0.5]{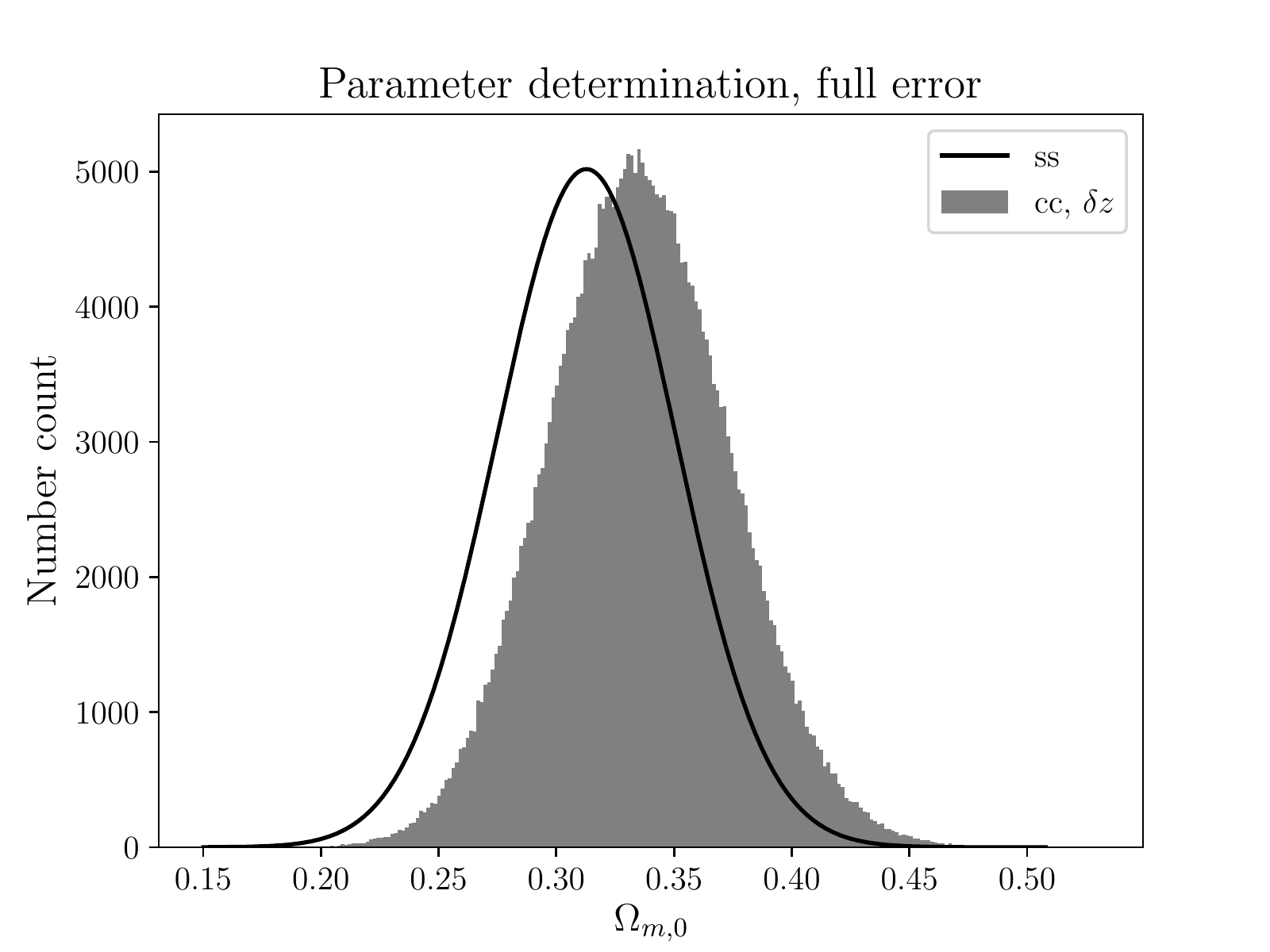}
	\caption{Distribution of $\Omega_{m}$ in MCMC chains obtained with emcee \cite{emcee} using the full error when generating mock data. The distribution marked ``ss'' represents results obtained with the standard siren data, i.e. by wrongfully employing equation \ref{eq:dipole_flat} on mock dipole luminosity distance data from DECIGO. The distribution marked ``cc, $\delta z$'' represents mock data yielding the true $H(z)$, such as cosmic chronometers or redshift drift data.}
	\label{fig:omega_full}
\end{figure}
Based on the above logic, the mock data is used to constrain the flat $\Lambda$CDM model. Assuming that $H_0$ is measured by other means (and ignoring any errors there may be on its measurement), this means that the model only  has one adjustable parameter, namely $\Omega_{m} = 1-\Omega_{\Lambda}$. Using emcee \cite{emcee}, mock data based on a curved $\Lambda$CDM model with $\Omega_{m} = 0.35$, $\Omega_{\Lambda} = 0.7$ and $H_0 = 70$km/s/Mpc (i.e. the data to the left in figure \ref{fig:mock}) is used to constrain $\Omega_{m}$. The two data sets -- one based on wrongly employing equation \ref{eq:dipole_flat} to dipole luminosity distance data and one set based on true $H(z)$ values -- are treated separately.\footnote{At this point, it should perhaps also be emphasized that the two mock data sets are still in principle inconsistent if parameter estimates are based on a curved $\Lambda$CDM model, but in this case the data lacks constraining power and the IOI does not indicate inconsistency. This can be remedied by adding more data points in which case the IOI can be driven to higher values. This interpretation is somewhat contrived though; presumably one would not use the data to do a parameter determination permitting curvature without also remembering to include the effect of curvature in the data reduction process, in which case the dipole luminosity distance data does not generate $H(z)$ data as equation \ref{eq:dipole_FLRW} shows.}
\newline\indent
\begin{table*}[]
	\centering
	\begin{adjustbox}{width=1\textwidth}
	\begin{tabular}{c c c c c c c c c c c}
		\hline\hline
	     IOI: &6.8 & 6.5& 1.2 & 6.7 &2.8 & 0.63&0.93 &0.009 & 4.5 & 1.5 \\
	     $\Omega_{m}$ (ss): & $0.338^{+0.0035}_ {-0.0035}$ & $ 0.337^{+0.0035}_ {-0.0035} $& $0.332^{+0.0035}_{-0.0034}$ & $0.335^{+0.0035}_{-0.0035}$ & $0.339^{+0.0035}_ {-0.0035}$ & $0.329^{+0.0034}_{-0.0035}$ & $0.333^{+0.0035}_{-0.0035}$ & $0.329^{+0.0035}_{-0.0035}$ & $0.342^{+0.0035}_{-0.0035}$ & $0.326^{+0.0034}_{-0.0034}$ \\
	     $\Omega_{m}$ (cc, $\delta z$): & $0.321^{+0.0034}_{-0.0034}$ & $0.319^{+0.0034}_{-0.0034}$ & $0.324^{+0.0034}_{-0.0034}$ &$0.318^{+0.0034}_{-0.0034}$& $0.328^{+0.0034}_{-0.0034}$& $0.324^{+0.0034}_{-0.0034}$ & $0.326^{+0.0034}_{0.0034}$ & $0.328^{+0.0034}_{-0.0034}$& $0.327^{+0.0034}_{-0.0034}$& $0.334^{+0.0034}_{-0.0035}$\\
		\hline
	\end{tabular}
	\end{adjustbox}
	\caption{IOI (index of inconsistency) and mean and error on $\Omega_{m}$ using different mock data sets, all with an error of 1.5\% and redshift bin width of 0.1, corresponding to 10 data points in the redshift interval $z\in[0,1]$.}
	\label{table:IOI_10}
\end{table*}
\begin{table*}[]
	\centering
	\begin{tabular}{c c c c c}
		\hline\hline
		Range: & IOI$<1$ & $1<\rm{IOI}<2.5$ & $2.5<\rm IOI<5$ & $\rm IOI>5$ \\
		\hline
		Inconsistency: & negligible & weak & moderate & strong\\
		\hline
	\end{tabular}
	\caption{Jeffrey's scale for interpreting IOI. Based on figure III of \cite{discordance}.}
	\label{table:jeffrey}
\end{table*}
Figure \ref{fig:omega_15} shows the distribution of $\Omega_{m}$ obtained from the MCMC chains. Since both data sets are analyzed assuming a {\em flat} $\Lambda$CDM model, neither of the estimated parameter values are correct. This is apparent already from the figure but is also shown in table \ref{table:IOI_10} which presents mean values and 1-$\sigma$ errors\footnote{The mean was computed as the 50th percentile of the data while the standard deviations were computed through the 15.9th and 84.1st percentiles. This corresponds to the usual definition of mean and standard deviation if the data is Gaussian which figure \ref{fig:omega_15} reveals is a good approximation.} of $\Omega_{m}$ as well as the computed IOI based on data generated with 10 different random seeds. Since only a single parameter is constrained by the data set, the IOI is simply computed according to
\begin{align}
 \text{IOI} = \frac{1}{2}\frac{\delta^2}{\sigma_{(1)}^2+\sigma_{(2)}^2},
\end{align}
where $\delta$ is the difference between the mean values of $\Omega_{m}$ and $\sigma_{(i)}$ is its standard deviation obtained with the two data sets.
\newline\newline
\begin{table*}[ht]
	\centering
	\begin{adjustbox}{width=1\textwidth}
		\begin{tabular}{c c c c c c c c c c c}
			\hline\hline
			IOI: & 3.1 & 2.8& 1.2& 2.1&11.2& 4.6&2.7& 12.3& 10.5& 10.8\\ 
			$\Omega_{m}$ (ss): & $0.337^{+0.0024}_{-0.0024}$ & $0.336^{+0.0024}_{-0.0024}$& $0.334^{+0.0024}_{-0.0024}$  & $0.334^{+0.0024}_{-0.0024}$ &$0.336^{+0.0024}_{-0.0024}$& $0.336^{+0.0024}_{-0.0024}$ & $0.337^{+0.0024}_{-0.0024}$ & $0.347^{+0.0024}_{-0.0024}$& $0.342^{+0.0024}_{-0.0024}$& $0.338^{+0.0023}_{-0.0024}$\\
			$\Omega_{m}$ (cc, $\delta z$): & $ 0.328^{+0.0023}_{-0.0023}$&$0.328^{+0.0023}_{-0.0023}$& $0.329^{+0.0024}_{-0.0024}$& $ 0.328^{+0.0023}_{-0.0023}$&$0.320^{+0.0024}_{-0.0023}$&$0.326^{+0.0023}_{-0.0023}$& $0.329^{+0.0024}_{-0.0023}$& $0.330^{+0.0023}_{-0.0024}$ & $0.327^{+0.0023}_{-0.0024}$ & $0.323^{+0.0023}_{-0.0023}$\\
			\hline
		\end{tabular}
	\end{adjustbox}
	\caption{IOI (index of inconsistency) and mean and error on $\Omega_{m}$ using different mock data sets, all with an error of 1.5\% and redshift bin width of 0.05, corresponding to 20 data points in the redshift interval $z\in[0,1]$.}
	\label{table:IOI_20}
\end{table*}
The first entry in table \ref{table:IOI_10} represents the distributions depicted in figure \ref{fig:omega_15}. As seen, the IOI of this data fit it 6.8. As suggested in \cite{discordance}, this number should be compared to Jeffrey's scale summarized in table \ref{table:jeffrey} based on table III in \cite{discordance}. From the table it is seen that IOI$>5$ should be interpreted as a strong indication of inconsistency. However, as is also seen from table \ref{table:IOI_10}, using different random seeds when generating the mock data leads to different values of IOI because the data is so sparse. With 10 different data sets (shown in the table), the IOI varies between 0.63 and 6.8 -- i.e. depending on the specific random generation of the data set, the IOI ranges from indicating ``negligible inconsistency'' to ``strong inconsistency'' and only half the data sets indicate moderate or strong inconsistency. To illustrate that the problem is that only 10 data points are used (with a precision of $1.5\%$), the analysis has been redone using 10 different random seeds to generate 20 data points with a precision of $1.5\%$. The results from the MCMC analysis on these mock data sets are shown in table \ref{table:IOI_20}. Now, the 10 different parameter determinations agree better with generally quite high values of ITI. There is still some variation in the IOI values though, but only two of the IOI values are below 2.5, while the remaining 8 values indicate either strong or moderate discordance.
\begin{table}[]
	\centering
	\begin{tabular}{c c c c}
		\hline\hline
		Error &\,\,\ $\Omega_{m}$ from ss \,\,\ & $\Omega_{m}$ from cc, $\delta z$ \,\,\ & IOI \\
		\hline\\
		1.5\% & $0.338^{+0.00347}_ {-0.00348}$ & $0.321^{+0.00338}_{-0.00341}$ & 6.8 \\
		\\
		full & $0.313^{+0.0376}_{-0.0383}$  & $0.335^{+0.0377}_{-0.0390}$ & 0.082\\
		&  &  &\\
		\hline
	\end{tabular}
	\caption{Parameter determination and IOI (index of inconsistency) from mock data generated with an error of $1.5\%$ or using the full expression for the error.}
	\label{table:IOI}
\end{table}
\newline\indent
The expectation is, however, not that DECIGO or other planned gravitational wave detectors will yield this number -- 20 -- of effective/binned data points with such high precision (at least not within a 10-year survey). Indeed, the precision of 1.5\% for 10 (binned) data points is already quite optimistic. Using the full error described in section \ref{subsub:generating_mock} yields the distribution shown in figure \ref{fig:omega_full} (using the same random seed as that used for generating mock data for figure \ref{fig:omega_15}). The resulting IOI and $\Omega_{m}$ estimates are shown in table \ref{table:IOI} with a comparison to results from one of the data sets with a $1.5\%$ precision. The most important point with this comparison is that the constraining power of the data set with the full error is significantly lower than when the error is $1.5\%$. With the full error, around a factor of $10^4$ more data points are required to reduce the error (which scales as $\propto 1/\sqrt{N}$, with $N$ the number of data points) enough to be able to obtain reliable IOI estimates that consistently reflect discordance between the two types of data sets (standard siren/luminosity distance dipole versus true measurements with e.g. cosmic chronometers and redshift drift). Detecting such a high number of merger events does currently not seem realistic (within a 10-year period). Thus, techniques to remove the lensing error {\em must} be introduced in order for the IOI to signal a discordance between $d_L^1$-based and true measurements of $H(z)$. Even if this can be achieved, a reliable IOI measure will require better/more data than what is currently expected. Alternatively, $H(z)$ data from cosmic chronometers or redshift drift need to become more precise and ample than what was (somewhat ad hoc-ly) assumed here.

\section{Discussion and conclusions}
An expression for the dipole of the luminosity distance was derived under the assumptions of spatial statistical homogeneity and isotropy but without requiring the Universe to be FLRW and/or flat on large scales. From this derivation, it is clear that the dipole of the luminosity distance, $d_L^1$, does not in general directly yield a measure of the large scale expansion rate -- even in cosmologies which are statistically homogeneous and isotropic and assuming a Copernican observer. If the Universe is not well described by a single flat FLRW model, this fact will in principle lead to a discordance between the ``$H(z)$'' measurements obtained by using the usual flat FLRW relation between $H(z)$ and $d_L^1(z)$ on the one hand, and $H(z)$ obtained from e.g. cosmic chronometers and redshift drift measurements on the other hand. Using concrete examples of curved FLRW models and backreaction models based on the scaling solution, it was studied if estimates of the large scale expansion rate obtained with DECIGO measurements of $d_L^1$ will be precise enough to reveal a deviation from the true large scale expansion rate. In a qualitative analysis it was first illustrated that the $H(z)$ measured with DECIGO is in principle distinguishable from the true $H(z)$ if the error is optimistically set to 1.5 \% and the curvature is above 1 \%. For the studied scaling solutions, the backreaction effect instead has to be about 10 \% before it becomes detectable using two particular values of the scaling parameter $n_D$ of the scaling solutions. However, the value of $n_D$ has a large impact on the possibility of detecting the difference. Since the scaling solutions are not particularly well motivated on physical grounds, this quantification can thus not be considered too reliable. The results nonetheless show that cosmic backreaction {\em in principle} shows up as a signal in the $H(z)$ measurements based on the dipole luminosity distance because backreaction affects the expression of $d_L^1$ so that it is no longer given by the usual equation \ref{eq:dipole_flat}. Whether or not the signal will be detectable through the data discordance discussed here has to be determined on a model-by-model basis.
\newline\indent
A quantitative analysis was carried out with 10-year mock DECIGO data generated with an FLRW model with $\Omega_{m}$ increased to 0.35 to generate a 5\% curvature. It was found that even with the optimistic $1.5\%$ error on the data, it is not possible to reliably use the index of inconsistency, IOI, to determine that the $H(z)$ measurement from $d_L^1$ is in discordance with true measurements of $H(z)$. This is likely because the high precision of $1.5\%$ requires binning of data which means that only 10 effective data points are left. If the precision of $1.5\%$ is kept and the number of data points is increased to 20, 8 out of 10 computed IOI values indicate moderate or strong inconsistency between $\Omega_{m}$ estimates based on the two different types of data sets -- i.e. the data set corresponding to $H(z)$ wrongly being obtained using the flat FLRW relation between $d_L^1$ and $H(z)$ versus the correct $H(z)$ obtained using mock data from e.g. cosmic chronometers or redshift drift.
\newline\indent
A crucial point that should be stressed is that the possibility of detecting signatures of curvature and/or cosmic backreaction with measurements of the dipole of the luminosity distance depends not only on the precision of the measurement of the dipole itself. It also depends on the precision with which the true Hubble parameter can be measured. Measuring this currently only seems to be possible using cosmic chronometers and the errors on cosmic chronometers data are quite large -- often several tens of percent. But the method of using cosmic chronometers is also still fairly new, so there is hope that the method will lead to more precise determinations of $H(z)$ in the future. Future redshift drift measurements will also be able to directly probe the Hubble parameter in curved FLRW spacetimes, but this does not seem to be the case in more general spacetimes such as those with significant cosmic backreaction. Since the future prospects of obtaining true measurements of $H(z)$ with these methods is currently unclear, mock data was here generated using the same distribution and error on true $H(z)$ measurements as on $d_L^1$-based estimates of $H(z)$.

\section{Acknowledgments}
The author is funded by the Carlsberg foundation. The author thanks the anonymous referee for suggestions that have significantly developed and improved the manuscript. The author also thanks Asta Heinesen for comments on the manuscript and Jing-Zhao Qi for correspondence and code sharing regarding the construction of mock DECIGO data.
\newline\indent
The redshift distribution of DECIGO data was computed using the publicly available code of \cite{z_sample} and binning of mock data was carried out using PyAstronomy \cite{Pyastronomy}. 

\end{document}